\documentclass[a4paper]{article}  % must use LaTeX2e

\def\comment#1{} % invisible remark in manuscript
\def\journalfont{\rm}         % this allows redefinition of the font later
\def\jou#1{{\journalfont #1\ }}
\def\joudef#1#2{\def #1{\jou{\ignorespaces #2}}}

\joudef{\aaa}    { Astron.\ Astrophys.}
\joudef{\aip}    { Adv.\ Phys.}
\joudef{\adm}    { adv.\ math.}
\joudef{\aihpa}  { Ann.\ Inst.\ H.\ Poincar\'e A}
\joudef{\am}     { Ann.\ Math.}
\joudef{\apny}   { Ann.\ Phys.\ (N.Y.)}
\joudef{\arns}   { Ann.\ Rev.\ Nucl.\ Sci.}
\joudef{\arnps}  { Annu.\ Rev.\ Part.\ Sci.}
\joudef{\apj}    { Astrophys.\ J.}
\joudef{\cjp}    { Can.\ J.\ Phys.}
\joudef{\cmp}    { Commun.\ Math.\ Phys.}
\joudef{\cqg}    { Class.\ Quantum Grav.}
\joudef{\grg}    { Gen.\ Rel.\ Grav.}
\joudef{\gp}     { Geom.\ and Phys.}
\joudef{\ijmpd}  { Int.\ J.\ Mod.\ Phys.\ D}
\joudef{\ijtp}   { Int.\ J.\ Theor.\ Phys.}
\joudef{\invm}   { Invent.\ Math.}
\joudef{\jm}     { J.\ Math.}
\joudef{\jmaa}   { J.\ Math.\ Anal.\ Appl.}
\joudef{\jmp}    { J.\ Math.\ Phys.}
\joudef{\jpa}    { J.\ Phys.\ A}
\joudef{\mnras}  { Mon.\ Not.\ R.\ Ast.\ Soc.}
\joudef{\mpla}   { Mod.\ Phys.\ Lett.\ A} 
\joudef{\nature} { Nature}
\joudef{\nc}     { Nuovo Cim.}
\joudef{\ncb}    { Nuovo Cim. B}
\joudef{\npa}    { Nuc.\ Phys.\ A}
\joudef{\npb}    { Nuc.\ Phys.\ B}
\joudef{\ph}     { Physica}
\joudef{\pla}    { Phys.\ Lett. A}
\joudef{\plb}    { Phys.\ Lett. B}
\joudef{\pr}     { Phys.\ Rev.}
\joudef{\prd}    { Phys.\ Rev.\ D}
\joudef{\prep}   { Phys.\ Rep.}
\joudef{\prl}    { Phys.\ Rev.\ Lett.}
\joudef{\prsla}  { Proc.\ Roy.\ Soc.\ Lond.\ A}
\joudef{\ptp}    { Prog.\ Theor.\ Phys.}
\joudef{\ptps}   { Prog.\ Theor.\ Phys.\ Suppl.}
\joudef\rmp      { Rev.\ Mod.\ Phys.}
\joudef\spj      { Sov.\ Phys.\ JETP}
\catcode`@=11

\def\eqalign#1{\null\,\vcenter{\openup\jot\m@th
  \ialign{\strut\hfil$\displaystyle{##}$&$\displaystyle{{}##}$\hfil
      \crcr#1\crcr}}\,}
\def\meqalign#1{\null\,\vcenter{\openup\jot\m@th
  \ialign{\strut\hfil$\displaystyle{##}$&&$\displaystyle{{}##}$\hfil
      \crcr#1\crcr}}\,}
\catcode`@=12   % back to nonletter category 
\newdimen\arrayruleHwidth
\makeatletter
\def\Hline{\noalign{\ifnum0=`}\fi\hrule \@height \arrayruleHwidth
  \futurelet \@tempa\@xhline}
\makeatother
\newcommand\thickbaselines{\baselineskip=20pt\lineskip=3pt\lineskiplimit=3pt}
\catcode`@=11
\def\cases#1{\left\{\,\vcenter{\thicknormalbaselines\m@th
             \ialign{$##\hfil$&\quad##\hfil\crcr#1\crcr}}\right.}
\def\matrix#1{\null\,\vcenter{\thickbaselines\m@th
    \ialign{\hfil$##$\hfil&&\quad\hfil$##$\hfil\crcr
      \mathstrut\crcr\noalign{\kern-\baselineskip}
      #1\crcr\mathstrut\crcr\noalign{\kern-\baselineskip}}}\,} 
\catcode`@=12   % back to nonletter category

\newcommand{\eprint}{\textsf} % style for electronic references
\newcommand\be{\begin{equation}} \newcommand\ee{\end{equation}} % numbered
\newcommand\bd{\begin{displaymath}}\newcommand\ed{\end{displaymath}}%unnumb'd

\newcommand\sinc{\mathop{\rm sinc}\nolimits}
\newcommand{\re}{\Re{e}}

\newcommand\ts\textstyle

\def\undersim#1{\mathop{\vtop{\ialign{##\crcr
     $\hfil\displaystyle{#1}\hfil$\crcr\noalign
     {\kern1pt\nointerlineskip}\hbox{$\hfil\sim\hfil$}\crcr
     \noalign{\kern1pt}}}}}

\def\eg{{\it e.g.\ }}  \def\ie{{\it i.e.\ }}
\def\cf{{\it cf.\ }}

\usepackage{amsmath,amssymb,bbm} 
\usepackage{graphicx}

\newcommand{\ncd}{\newcommand} 
\ncd{\Dtwo}{\mathcal{D}}
\ncd{\Ttwo}{\mathcal{T}}
\ncd{\lagomdot}{{\mbox{\large$\cdot$}}}
\ncd{\dotprime}{^{(\lagomdot}{}'{}^)}
\ncd{\nms}{\negmedspace} 
\ncd{\nts}{\negthickspace} 
\ncd{\mcl}[1]{\mathcal{#1}} 
\ncd{\beq} {\begin{equation}} 
\ncd{\eeq} {\end{equation}} 
\ncd{\BE} {\begin{eqnarray}} 
\ncd{\EE} {\end{eqnarray}} 
\ncd{\rarr} {\rightarrow} 
\ncd{\larr} {\leftarrow} 
\ncd{\lrarr} {\leftrightarrow} 
\ncd{\lbeq}[1]  {\label{eq: #1}} 
\ncd{\refeq}[1] {(\ref{eq: #1})} 
\ncd{\mrm}    {\mathrm} 
\ncd{\nn}{\nonumber} 
\ncd{\mbf}[1] {{\mathbf #1}} 
\ncd\T{\frac{1}{2}h^{\mu\nu}p_\mu p_\nu} 
\ncd{\ms}{\mathstyle} 
\ncd{\ds}{\displaystyle} 
\ncd{\bmth}[1] {\mbox{\boldmath $#1$}} 
\ncd{\abs}[1] {|#1|} 
\ncd{\ubold}{\mathbf u}
\ncd{\Abold}{\mathbf A}
\ncd{\Bbold}{\mathbf B}
\ncd{\Mbold}{\mathbf M}
\ncd{\tsfrac}[2]{{\ts\frac{#1}{#2}}}
\ncd{\lagom}{\hspace{.6pt}}
\ncd{\muk}{k}
\ncd{\dumkonstant}{v_0}
\ncd{\tdelta}{{\tilde\delta}}
\ncd{\Q}{\Theta}

\newtoks\reportnoregister \newtoks\eprintnoregister
\newcommand{\reportnumber}[1]{\reportnoregister={#1}}
\newcommand{\eprintnumber}[1]{\eprintnoregister={#1}}

\reportnumber{\mbox{}} % empty registers to start with, needed if the
\eprintnumber{\mbox{}} % \reportnumber (or \eprintnumber) command isn't used

\newcommand{\reportid}{
   \begin{minipage}{17cm}\vspace{-7.2cm}
     \begin{flushright}
      {\normalsize \the\reportnoregister \\[-.2cm]
            \eprint{\the\eprintnoregister}}\vspace{0.2cm}
     \end{flushright}
   \end{minipage}\hspace{-17cm} }

\catcode`@=11   % make @ act like letter temporarily
\def\title#1{\gdef\@title{\reportid#1}}
\catcode`@=12   % back to nonletter category

\begin{document} 

\reportnumber{USITP 2004-2}
\eprintnumber{gr-qc/0401115}

\title{Elastic Stars in General Relativity:\\ III. Stiff ultrarigid exact solutions}
  \author{Max Karlovini\footnote{E-mail: \eprint{max@physics.muni.cz}, Present address: Department of Theoretical Physics and Astrophysics, Faculty of Science, Masaryk University, Kotl\'a\u{r}sk\'a 2, 611 37 Brno, Czech Republic},
  Lars Samuelsson\footnote{E-mail: \eprint{larsam@physto.se}} \\[10pt]
  {\small Department of Physics, Stockholm University}  \\
  {\small AlbaNova University center, 106 91 Stockholm, Sweden} }
\date{} \maketitle

%\vspace{2cm}

\begin{abstract}{\normalsize
    We present an equation of state for elastic matter which allows
    for purely longitudinal elastic waves in all propagation
    directions, not just principal directions. The speed of these
    waves is equal to the speed of light whereas the transversal type
    speeds are also very high, comparable to but always strictly less
    than that of light. Clearly such an equation of state does not
    give a reasonable matter description for the crust of a neutron
    star, but it does provide a nice causal toy model for an extremely
    rigid phase in a neutron star core, should such a phase exist.
    Another reason for focusing on this particular equation of state
    is simply that it leads to a very simple recipe for finding
    stationary rigid motion exact solutions to the Einstein equations.
    In fact, we show that a very large class of stationary spacetimes
    with constant Ricci scalar can be interpreted as rigid motion
    solutions with this matter source. We use the recipe to derive a
    static spherically symmetric exact solution with constant energy
    density, regular centre and finite radius, having a nontrivial
    parameter that can be varied to yield a mass-radius curve from
    which stability can be read off.  It turns out that the solution
    is stable down to a tenuity $R/M$ slightly less than $3$. The
    result of this static approach to stability is confirmed by a
    numerical determination of the fundamental radial oscillation mode
    frequency. We also present another solution with outwards
    decreasing energy density.  Unfortunately, this solution only has
    a trivial scaling parameter and is found to be unstable.}
\end{abstract}
\vspace{.5cm}
\centerline{\bigskip\noindent PACS: 04.40.Dg, 97.10.Cv, 97.60.Jd}

%%%%%%%%%%%%%%%%%%%%%%%%%%%%%%%%%%%%%%%%%%%%%%%%%%%%%%%%%%%%%%%%%%%%%%%%%
\section{Introduction}
There are few exact solutions to the Einstein equations that can be
considered stellar models. The most well-known solution is the
interior Schwarzschild solution which can be thought as a limiting
case for perfect fluids since it satisfies the Buchdahl inequality
$R/M \leq 9/4$ as an equality when the central pressure is taken to
infinity. However, the constant energy density of this solution makes
it unphysical, at least when interpreting it as an adiabatic perfect
fluid since the speed of sound is then infinite. Although there are
other known perfect fluid solutions, only a few of them satisfy the
basic criteria of physicality, \cf \cite{dl:physsss} for a
comprehensive classification of known exact solutions. Several workers
have also considered equilibrium models with pressure anisotropy, at
least since the paper \cite{bl:aniso}. In many cases an ``exact
solution'' is obtained through some ansatz which does not correspond
to a particular matter description. Elastic matter, on the other hand,
gives rise to anisotropic pressures in a natural way and should
moreover be relevant for modelling of \eg neutron stars since they are
believed to have solid crusts and, perhaps, solid cores. The general
relativistic theory of elasticity has never been a very hot field of
research, but over the years it has generated quite a number of
papers, most of which concern the theoretical formulation of the
theory. Despite of this, the times when the theory -- in its full
nonlinear regime -- has found its way into astrophysical applications
are scarce.
%be counted on the fingers on the left hand of a blind lumberjack.

In a recent paper by the present authors\cite{ks:relasticityI}
(hereinafter paper I), we reconsidered the subject and applied it to
static spherically symmetric (SSS) configurations. In particular we
constructed elastic stellar models numerically for a specific equation
of state. In a subsequent paper\cite{ksz:stability} (paper II) we
showed how to analyse the radial oscillations of elastic stars and
found, in particular, that the models constructed in paper I are
stable up to the first maximum of the mass-radius curves, as one would
expect. Although one will probably always be forced to resort to
numerical methods when one wants to model a real star, it would
nevertheless be useful to have a simple exact solution that is not too
unphysical. Such a solution would serve as a convenient tool for
investigating qualitative features and for testing numerical codes. In
this context it could be useful to list some criteria that an exact
SSS solution should satisfy in order to be of astrophysical interest.
In our humble opinion, this list should minimally include
\begin{itemize}
  \item The centre should be regular (elementary flat), implying
  pressure isotropy there.
  
  \item The energy density $\rho$ should be positive at the centre 
        and everywhere non-negative.
  
  \item The isotropic central pressure $p_c$ should be positive.
  
  \item The surface of the model, where the radial pressure $p_r$ by
    necessity is zero, should be at a finite Schwarzschild radius
    $r=R$. Usually, although perhaps not necessarily, there are no
    other zeros of $p_r$ at smaller radii, implying that $p_r>0$
    everywhere inside the surface of the star.
  
  \item The three elastic wave modes associated with any propagation
    direction should have speeds $v$ obeying $0\leq v^2 \leq 1$ in
    geometrical units. In other words the equation of state, at least
    in the range it is used, should be microstable ($v^2\geq 0$) and
    causal ($v^2\leq 1$).

  \item The model should be stable against radial perturbations. 
\end{itemize}
Exact solutions with elastic matter sources have previously been
constructed by Magli and Kijowski\cite{mk:nonrot}. These solutions do
not obey all of the first four criteria however, while to our
knowledge it is undetermined whether they obey the last two. On the
other hand it should be stressed that a solution may fulfill the
criteria only when matched to some other matter region instead of used
as a stellar model in its own right. In this paper we will present a
family of solutions having a branch on which all of the criteria are
satisfied. The contents and structure of the present paper are as
follows;

In section \ref{sec:eos} we introduce and discuss the properties of a
particular equation of state, which we refer to as the \emph{stiff
  ultrarigid equation of state} (SUREOS). The most important features
of this equation of state are listed below.
\begin{itemize}
\item The SUREOS can be split into a conformally invariant part and a
  vacuum energy density $B$, leading to the relation
  $\rho=p_1+p_2+p_3+4B$ for the energy density and principal pressures
  which in turn implies that the spacetime Ricci scalar is constant
  and equal to $4\kappa B$, with $\kappa=8\pi$ in geometrical units.
  In this sense the SUREOS is similar to the MIT bag (perfect fluid)
  equation of state $\rho=3p+4B$ describing non-interacting strange
  quark matter in the limit of zero quark
  mass\cite{fj:strange}. Furthermore, the energy density and principal
  pressures are required to satisfy the inequality $\rho > p_\mu + 2B$
  for all $\mu$.
\item The SUREOS is \emph{stiff} in the sense that the speed of sound
  for longitudinal elastic waves propagating in principal directions
  is marginally causal, \ie equal to the speed of light. In fact the
  SUREOS allows for one purely longitudinal wave mode with marginally
  causal sound speed in any propagation direction. In this sense the
  SUREOS is similar to the stiff perfect fluid equation of state
  $\rho=p+\rho_0$ proposed by Zel'dovich\cite{z:stiff}.
\item The SUREOS is \emph{ultrarigid} in the sense that it has a very
  high shear modulus, of the same order as the energy density. This
  results in very high propagation speeds for the remaining two
  elastic wave modes that are purely transversal in principal
  directions. In an unsheared state these speeds are $1/\sqrt2$ times
  the speed of light. In general they are always real, positive and
  strictly less than the speed of light. 
\end{itemize}

In section \ref{sec:stationary} we give a recipe for obtaining
stationary rigid motion solutions to the Einstein equations with
SUREOS elastic matter source. The recipe simply amounts to making the
spacetime Ricci scalar take a constant value to be identified with
$4\kappa B$ and, in addition, to ensure that the inequalities
$\rho>p_\mu+2B$ are satisfied by the eigenvalues of the resulting
stress-energy tensor. As soon as this has been achieved one has an
exact solution to the Einstein equations with SUREOS elastic matter
source, with respect to a material space metric that can be explicitly
constructed according to a simple algebraic formula. Clearly one can
take the view that this method of letting the solution determine the
material space metric, rather than the other way around, is cheating.
Indeed, should one want to specify the material space metric by e.g.\ 
taking it to be flat, then the method should be of little use since
the resulting ``backwards'' constructed metric will in general be
curved. Although a flat metric is the correct choice when one wants to
describe a solid region without any lattice dislocations (\cf section
3 in paper I), such a desciption will only be appropriate throughout a
rather small solid region of \eg a neutron star. In particular, a neutron
star crust, having a thickness of the order 1 km, is bound to contain
many dislocations. Unless one wants to go into a detailed description
of the dislocations, their average effect over distances containing
many of them can instead be considered to be encoded into an
effective, generally curved, material space metric. In the case of
neutron star crusts this effective metric is in principle ``created''
at the moment when the outer region of a new born neutron star cools
down to the temperature at which solidification occurs, or at the
moment when an older/colder neutron star settles down after a star
quake (rupture of the crust). Analogous remarks apply to solid cores
in neutron stars, should such cores exist. When constructing an
equilibrium stellar model, the simplest choice of effective material
space metric is the one that results (usually only implicitly) from
assuming that the star is in an unsheared state, implying pressure
isotropy so that the star can alternatively be thought of as being
made up out of perfect fluid matter. In this case the rigidity of the
stellar matter only enters through a nonvanishing shear modulus when
considering stellar perturbations around the equilibrium model.
Whereas this approach is fine under many circumstances, it does not
allow for studies of how the perturbations are affected by pressure
anisotropies of the background. To test the effect of anisotropies one
can use moderately (and hence realistically) anisotropic equilibrium
models, such as the ones constructed numerically in paper I, or some
exact solution which is likely to have more extreme pressure profiles,
should it be possible to find one that obeys the basic list of
criteria set up above. This paper goes to show that this indeed is
possible.

In section \ref{sec:exactfreeB} we apply the general recipe of section
\ref{sec:stationary} to construct a spherically symmetric family of
solutions for which each member has constant energy density. As a
direct consequence of the constant energy density and the equation of
state, the radial and tangential pressure cannot be simultaneously
decreasing with increasing radius because of the linear relation
$\rho=p_r+2p_t+4B$. Instead it turns out that the radial pressure is
monotonically decreasing from the centre to the surface while the
tangential pressure is monotonically increasing at half the rate. For
each fixed value of the equation of state parameter $B$ the solution
family has one remaining free parameter which can be chosen to be the
(isotropic) central pressure, going between zero and infinity. For
$B=0$ the equation of state is scale invariant and the central
pressure is just a trivial scaling parameter. However, for $B>0$ (we
see $B<0$ as unphysical) the freely specifiable central pressure
results in a nontrivial mass-radius curve, exhibiting a mass maximum.
From the conclusions drawn in paper II this maximum should be the
point where instability sets in. More precisely, since the only mass
extremum is for the maximum mass solution, the solutions with lower
central pressure should be stable whereas those with higher central
pressure should have exactly one unstable mode of radial oscillations,
given that it is known that the models are stable in the limit of zero
central pressure. This behaviour is exactly what a numerical
determination of the first two frequencies of radial oscillations
shows. The main motivation for this paper is the fact that the
solutions on the stable branch of the mass-radius curve satisfy
\emph{all} items on the above list of criteria for physicality of
exact spherically symmetric solutions. Due to the rather extreme
equation of state and the nature of the energy density and pressure
profiles, we make no claims that this solution family provides an
accurate description of real stars. However, solid neutron star cores
having very high shear moduli up to the same order of magnitude as the
pressure have actually been discussed in the literature, especially in
the first half of the 70s (\cf Haensel\cite{haensel:solid} for a
discussion, historical notes and further references). Should such
cores exist, despite the fact that they have not been theoretically
favoured in later years, the SUREOS would be a very relevant toy model
equation of state. Moreover, we wish to emphasize that in contrast to
the interior Schwarzschild models which also have constant energy
density, the elastic matter models presented here have causal speeds
of wave propagation.  The reason why causality is compatible with
constant energy density in the latter case is because the energy
density consists of two parts; one purely compressional part which is
outwards decreasing and one shearing part which is outwards increasing
due to increase of pressure anisotropy.  Clearly then, a constant
energy density can be obtained if these two effects exactly cancel
each other out, which is precisely what happens here.

In section \ref{sec:exactzeroB} we present another spherically
symmetric solution with $B=0$ for which the energy density as well as
the radial and tangential pressures all are outwards monotonically
decreasing, which in that sense makes it a more realistic model of a
star. The solution turns out to be unstable, however, which should not
be too surprising considering that a $B=0$ model with finite central
pressure can be viewed as a rescaled version of a $B\neq 0$ model with
infinite central pressure. Because of its instability we do not
discuss this solution in great detail, but it should be interesting to
generalise it to $B>0$ since we then expect there to be a stable
branch of solutions anologous to the stable branch of the constant
energy density family. The reason why there exists more than one
one-parameter solution for a given equation of state and central
pressure is due to the freedom of choosing the material space metric
in more than one way. Clearly it is this freedom that makes the simple
recipe for generating SUREOS solutions possible. Even more interesting
would be to use the recipe for finding rigidly rotating solutions, but
that is beyond the scope of the present paper.

The notation and conventions will follow papers I and II.

%%%%%%%%%%%%%%%%%%%%%%%%%%%%%%%%%%%%%%%%%%%%%%%%%%%%%%%%%%%%%%%%%%%%%%%%%
\section{The stiff ultrarigid equation of state (SUREOS)}\label{sec:eos}
%%%%%%%%%%%%%%%%%%%%%%%%%%%%%%%%%%%%%%%%%%%%%%%%%%%%%%%%%%%%%%%%%%%%%%%%%
As shown in paper I, given an elastic equation of state of the general
type associated with a material space metric,
\begin{equation}\lbeq{eosgen}
  \rho = \rho(n_1,n_2,n_3),
\end{equation}
where the $n_\mu$'s are the principal linear particle densities as
defined in paper I, the speed of a longitudinal elastic wave
propagating in the direction of a unit principal vector $e_\mu^{\;a}$
is given by
\begin{equation}\lbeq{vlong2}
  v_{\mu||}^{\;2} = \frac{\beta_\mu}{\rho+p_\mu}, \quad \beta_\mu = n_\mu\frac{\partial p_\mu}{\partial n_\mu}. 
\end{equation}
Using that the principal pressure $p_\mu$ is defined through the relation
\begin{equation}
  \rho+p_\mu = n_\mu\frac{\partial\rho}{\partial n_\mu}, 
\end{equation}
we can rewrite eq.\ \refeq{vlong2} as
\begin{equation}
  v_{\mu||}^{\;2} = \frac{n_\mu^{\;2}\displaystyle\frac{\partial^2\rho}{\partial n_\mu^{\;2}}}{n_\mu\displaystyle\frac{\partial\rho}{\partial n_\mu}}.
\end{equation}
Setting $v_{\mu||}$ to a constant, say $c_{||}$, it follows that
$\rho$ should satisfy the differential equation
\begin{equation}\lbeq{rhopde}
  n_\mu^{\;2}\frac{\partial^2\rho}{\partial n_\mu^{\;2}} = 
  c_{||}^{\;2}\,n_\mu\frac{\partial\rho}{\partial n_\mu}. 
\end{equation}
%Since we are only considering elastic materials that are intrinsically
%isotropic, the equation of state \refeq{eosgen} should be invariant
%under permutations of the $n_\mu$'s, implying that the solution to
%eq.\ \refeq{rhopde} is
%%%%brushed up language$$$$$
Since we are only considering elastic materials that are intrinsically
isotropic, the equation of state \refeq{eosgen} should be invariant
under permutations of the $n_\mu$'s. With this restriction the general
solution to eq.\ \refeq{rhopde} is
\begin{equation}
  \rho = C_1(n_1n_2n_3)^\alpha + C_2\left[(n_1n_2)^\alpha+(n_2n_3)^\alpha+(n_3n_1)^\alpha\right] + C_3(n_1^{\;\alpha}+n_2^{\;\alpha}+n_3^{\;\alpha}) + C_4, \quad \alpha = 1+c_{||}^{\;2}, 
\end{equation}
where $C_1$ - $C_4$ are constants. In this paper we focus on a
particular equation of state belonging to this class, namely the one
specified by setting $\alpha = 2$, $C_1 = C_3 = 0$ which we refer to
as the stiff ultrarigid equation of state (SUREOS) for reasons
explained in the introduction and below. Renaming the remaining two
constants as $C_2 = A$, $C_4 = B$ we find that the energy density and
principal pressures for the SUREOS are
\begin{align}\lbeq{eos}
  \rho &= A\left[(n_1n_2)^2+(n_2n_3)^2+(n_3n_1)^2\right] + B \\ \lbeq{pfunn}
  p_\mu &= n_\mu\frac{\partial\rho}{\partial n_\mu}-\rho =
  A\left[n_\mu^{\;2}(n_{\mu+1}^{\;2}+n_{\mu+2}^{\;2})-(n_{\mu+1}n_{\mu+2})^2\right]-B,
\end{align}
where we use the cyclic rule $\mu+3=\mu$.
This leads to the linear relation
\begin{equation}\lbeq{linrel}
  \rho = p_1 + p_2 + p_3 + 4B. 
\end{equation}
In an isotropic state, \ie when $n_1 = n_2 = n_3 = n^{1/3}$ leading to
$p_1 = p_2 = p_3 = p$, this relation mimics the MIT bag equation of
state $\rho = 3p + 4B$ with $B$ being the bag constant.  This directly
suggests that $B$ should be positive or at least non-negative, since
$B<0$ would lead to a negative energy density at zero isotropic
pressure. Now, while the speed of sound for the MIT bag equation of
state is $1/\sqrt3$ times the speed of light, the longitudinal speed
of sound in principal directions for the SUREOS is by construction
equal to the speed of light since we have set $\alpha = 2$. With the
longitudinal speed of sound being marginally causal in principal
directions, it is natural to wonder if it may become acausal in other
propagation directions. In general, neither longitudinal nor
transversal waves stay purely longitudinal/transversal when
continuously varying the propagation direction away from a principal
direction, but the SUREOS in fact allows for a purely longitudinally
polarized wave mode in \emph{all} propagation directions. Moreover,
the speed of these waves is direction independent and hence equal to
the speed of light.  These statements are proved in the appendix. What
about the remaining two polarization modes? As opposed to the
longitudinal mode their speeds generally depend on the polarization
direction. In principal directions the modes are purely transversal
with speeds $v_{\mu\perp\nu}$ given by
\begin{equation}\lbeq{transv2}
  v_{\mu\perp\nu}^{\;2} = 
  (\rho+p_\nu)^{-1}\frac{n_\nu^{\;2}(p_\nu-p_\mu)}{n_\nu^{\;2}-n_\mu^{\;2}} 
  = \frac{n_\sigma^{\;2}}{n_\sigma^{\;2}+n_\mu^{\;2}}, \quad 
   \mbox{$\sigma$ distinct from $\mu$ and $\nu$.}
\end{equation}
Here the indices $\mu$ and $\nu$ refer to two distinct principal unit
vectors $e_\mu^{\;a}$ and $e_\nu^{\;a}$, corresponding to the
propagation and polarization direction, respectively. Clearly, these
speeds are strictly within the range of microstability and causality,
obeying
\begin{equation}
  0< v_{\mu\perp\nu}^{\;2}< 1.
\end{equation}
Moreover, we prove in the appendix that the speeds of these modes are
less than the speed of light also in generic propagation directions. 

Although the SUREOS is neither of the quasi-Hookean type discussed in
paper I, nor of the closely related type introduced by Carter and
Quintana\cite{cq:elastica}, we can still define an unsheared energy
density $\check\rho$ and shear modulus $\check\mu$ that are functions
of the particle density $n = n_1n_2n_3$ only. This should in fact be
possible for any physically reasonable equation of state of the type
\refeq{eosgen}. The procedure is simply to consider the behaviour of
the equation of state for small shear, \ie when all linear particle
densities $n_\mu$ are close to the unsheared value $n^{1/3}$.  To this
end it turns out convenient to make the Misner type reparametrisation
\begin{equation}
\begin{split}\lbeq{reparam}
  n_1 &= n^{1/3}e^{2x_2} \\
  n_2 &= n^{1/3}e^{-x_2+\sqrt3\,x_1} \\
  n_3 &= n^{1/3}e^{-x_2-\sqrt3\,x_1}.
\end{split}
\end{equation}
Consider first the quasi-Hookean equation of state of paper I for
which the energy density is given by
\begin{equation}\lbeq{rhoqH}
  \rho_{\mathrm{q.H.}} = \check\rho + \check\mu s^2, 
\end{equation}
where $\check\rho$ and $\check\mu$ are functions of $n$ whereas $s^2$
is given by
\begin{align}
  s^2 &= \frac1{12}\left[ \left(\frac{n_1}{n_2}-\frac{n_2}{n_1}\right)^{\!\!2} 
  + \left(\frac{n_2}{n_3}-\frac{n_3}{n_2}\right)^{\!\!2} 
  + \left(\frac{n_3}{n_1}-\frac{n_1}{n_3}\right)^{\!\!2} \right] \\
  &= \frac13\left[\,\sinh^2{\!(\sqrt3\,x_1-3x_2)}+\sinh^2{\!(2\sqrt3\,x_1)}+\sinh^2{\!(\sqrt3\,x_1+3x_2)}\right]
\end{align}
Inserting the reparametrisation \refeq{reparam} into the equation of
state \refeq{eos} instead gives 
\begin{equation}\lbeq{rhostiff}
  \rho = An^{4/3}[e^{-4x_2} + e^{2(x_2-\sqrt3 x_1)} 
   + e^{2(x_2+\sqrt3 x_1)}] + B. 
\end{equation}
At fixed particle density $n$ this expression for $\rho$ can be
identified with a cyclic three-particle Toda lattice potential reduced
to the centre of mass system (\cf \cite{hietarinta:secinv}).  It
directly follows that the zero shear state \mbox{$x_1=x_2=0$} is the
only state that extremizes $\rho$ (and hence also the energy per
particle $\epsilon = \rho/n$) at fixed $n$. Moreover this extremum is
a minimum if $A$ is positive, which shall thus be assumed henceforth.
Taylor expanding the two energy densities \refeq{rhoqH} and
\refeq{rhostiff} in $x_1$ and $x_2$ gives
\begin{align}
  \rho_{\mathrm{q.H.}} &= \check\rho + 6\check\mu(x_1^{\;2}+x_2^{\;2}) + O(|x|^4) \\
  \rho &= 3An^{4/3} + B + 12An^{4/3}(x_1^{\;2}+x_2^{\;2}) + O(|x|^3). 
\end{align}
Hence we can identify the SUREOS as a quasi-Hookean equation of state
up to second order in $x_1$ and $x_2$, with unsheared energy density
and shear modulus given by
\begin{align}\lbeq{rhocheck}
  \check\rho &= 3An^{4/3} + B \\ \lbeq{mucheck}
  \check\mu &= 2An^{4/3}. 
\end{align}
Clearly the shear modulus is very large, of the same order as the
unsheared energy density unless $B\gg 3An^{4/3}$.
%\footnote{Note however that for a bcc 
%Coulomb lattice we have $\mu\approx0.1194(\frac{4\pi}{3})^{1/3}(Ze)^2n^{4/3}$. 
%[Strohmayer et. al., ApJ 375, 679-686 (1991)]
%Taking $Z\sim 1$ and $n\sim 10^{38}$ cm$^{-3}$ we find $3An^{4/3}\sim 10^{31}$ 
%dyn/cm$^2$, whereas 
%the bag constant is usually assumed to be $B^{1/4}\sim 150$ MeV which amounts to 
%$B\sim 10^{35}$ dyn/cm$^2$. It is however highly dubious to use a bcc lattice 
%calculation for a hypothetical quark lattice!}
As a consistency
check, let us now use eqs.\ \refeq{rhocheck} and \refeq{mucheck} to
calculate the longitudinal and transversal wave speeds $v_{||}$ and
$v_\perp$ with respect to an isotropic background. To do so we first
calculate the unsheared pressure and the bulk modulus: 
\begin{align}
  \check{p} &= n\frac{d\check\rho}{dn} - \check\rho = An^{4/3} - B 
   = \tsfrac13(\check\rho - 4B) \\
  \check\beta &= n\frac{d\check{p}}{dn} = \tsfrac43An^{4/3}.
\end{align}
Using the formulae of Carter\cite{carter:sound} (rederived in paper
I), we find that the isotropic wave speeds are
\begin{align}
  v_{||}^{\;2} &= \frac{\check\beta+\tsfrac43\check\mu}{\check\rho+\check{p}} 
  = \frac{d\check{p}}{d\check\rho}+\frac43\frac{\check\mu}{\check\rho+\check{p}} = 1 \\
  v_\perp^{\;2} &= \frac{\check\mu}{\check\rho+\check{p}} = \frac12. 
\end{align}
This is indeed consistent with our previous results, namely that the
longitudinal wave speed is always equal to the speed of light while
the principal transversal wave speeds are given by eq.\ 
\refeq{transv2}.

%%%%%%%%%%%%%%%%%%%%%%%%%%%%%%%%%%%%%%%%%%%%%%%%%%%%%%%%%%%%%%%%%%%%%%%%%%%
\section{Stationary solutions with rigid motion}\label{sec:stationary}
%%%%%%%%%%%%%%%%%%%%%%%%%%%%%%%%%%%%%%%%%%%%%%%%%%%%%%%%%%%%%%%%%%%%%%%%%%%
We here restrict our attention to stationary spacetimes, \ie
spacetimes admitting a timelike Killing vector $t^a$. We also assume
that the elastic matter source of Einstein's equations undergoes rigid
motion, by which we mean that the four-velocity $u^a$ is aligned with
$t^a$, \ie
\begin{equation}
  u^a = f^{-1}t^a, 
\end{equation}
for some scalar $f$. Since for elastic matter it always holds that
\begin{equation}
  h_a{}^bu^cT_{bc} = h_a{}^bu^c(\rho\,u_bu_c+p_{bc}) = 0, 
  \quad h_{ab} = u_au_b + g_{ab}, 
\end{equation}
it follows from Einstein's equations\cite{ksmhh:exact2ed} that the twist
of $t^a$ is locally the gradient of some scalar $\omega$,
\begin{equation}
  \omega_a = -\epsilon_{abcd}t^b\nabla^{c}t^{d} = \nabla_{\!a}\omega. 
\end{equation}
The four-dimensional physical spacetime $(M,g_{ab})$ is in this case
completely determined by the three-dimen\-sional structure
$(\Sigma,h_{ab},f,\omega)$, where $\Sigma$ is the three-manifold of
Killing orbits and where $h_{ab}$, $f$ and $\omega$ are fixed fields
on $\Sigma$. By ``fixed'' we mean that these fields are all anihilated
by $\mathcal{L}_t$, \ie constant along the Killing orbits.  Letting
$D_a$ denote the connection associated with $h_{ab}$, the Einstein
equations with stress-energy tensor $T_{ab} = \rho u_a u_b + p_{ab}$
can be written in three-dimensional form as\cite{ksmhh:exact2ed}
\begin{align}\lbeq{rho3d}
  \kappa\rho &= \tsfrac12{}^{(3)}\!R+\tsfrac34 f^{-4}D^a\omega\,D_a\omega \\ \lbeq{p3d}
  \kappa p_{ab} &= {}^{(3)}\!G_{ab} + f^{-1}[\,(D^cD_cf)h_{ab}-D_aD_bf\,] + \tsfrac14 f^{-4}[\,(D^c\omega D_c\omega)h_{ab} - 2D_a\omega D_b\omega\,]. 
\end{align}
Since the energy density $\rho$ and the three principal pressures
$p_\mu$ (the eigenvalues of $p^a{}_b$) all are constant along the
Killing flowlines there will exist at least one functional relation,
\begin{equation}\lbeq{Ffun}
  F(\rho,p_1,p_2,p_3)=0
\end{equation}
for any stationary spacetime of this type. It is not difficult to show
that if the relation \refeq{Ffun} can be written such that it is
invariant under permutations of the $p_\mu$'s, then the spacetime
represented by $h_{ab}$, $f$ and $\omega$ on $\Sigma$ can, under quite
general circumstances, be interpreted as a rigid motion solution to
the Einstein equation for at least \emph{some} elastic equation of
state $\rho=\rho(n_1,n_2,n_3)$ and an associated material space metric
$k_{AB}$. Explicitly, if the relation
$p_\mu=n_\mu\,\partial\rho/\partial n_\mu - \rho$ is inserted into
eq.\ \refeq{Ffun}, the latter becomes a partial differential equation
for the dependent variable $\rho(n_1,n_2,n_3)$. If the mentioned
permutation symmetry holds there will generically exist\footnote{We
  make no attempt to sort out under what precise conditions this will
  be true, since it is not our goal of this discussion.  Such an
  analysis would probably be nontrivial since eq.\ \refeq{Ffun} can in
  principle be a complicated nonlinear function of its arguments.},  at
least locally, a nonempty family of solutions $\rho(n_1,n_2,n_3)$ that
are invariant under permutations of the $n_\mu$'s and hence can be
interpreted as elastic equations of state of the type we focus on.
Picking one member of the family, each linear particle density $n_\mu$
becomes an implicit function of three independent combinations of
$\rho$ and the $p_\mu$'s, for instance by inverting the three
relations $p_\mu = p_\mu(n_1,n_2,n_3)$. This in turn determines the
tensor $k_{ab} = \sum_{\mu=1}^3n_\mu^{\;2}e_{\mu\,a}e_{\mu\,b}$ which
by this construction trivially satisfies the necessary and sufficient
conditions $u^ak_{ab}=0$ and $\mathcal{L}_u k_{ab} = 0$ for it to be
the pullback of a material space three-metric $k_{AB}$ from a
three-manifold that can be identified with $\Sigma$. Since $k_{AB}$
should be positive definite, one must finally require that
$n_\mu^{\;2}>0$ for all $\mu$. It should be stressed that this method
is in general of little practical use when it comes to finding exact
solutions, since the equation of state and the material space metric
will only be implicitly defined and cannot, except in special cases,
be expressed in closed analytic form.  However, one such special case
is provided by the SUREOS, \ie by eq.\ \refeq{eos}, for which eq.\ 
\refeq{Ffun} corresponds to the linear relation \refeq{linrel}. That
linear relation is identically satisfied for the SUREOS but in fact
also for a wider class of equations of state that we do not discuss
here. Taking thus the SUREOS to be the equation of state we can invert
the relation \refeq{pfunn} to find that the linear particle densities
are given by
\begin{equation}\lbeq{ninvert}
  n_\mu^{\;4} 
  = \frac{(\rho-p_{\mu+1}-2B)(\rho-p_{\mu+2}-2B)}{2A(\rho-p_\mu-2B)},
\end{equation}
where we have used eq.\ \refeq{linrel}. Since $\rho-p_\mu-2B$ are the
eigenvalues of $\Q^a{}_b = (\rho-2B)h^a{}_b-p^a{}_b$, the inverted
formulae \refeq{ninvert} implies that $k_{ab}$ can be calculated from
the stress-energy tensor via $\Q^a{}_{b}$ according to
\begin{equation}\lbeq{matmet}
  k_{ab} = \sum_{\mu=1}^3n_\mu^{\;2}\,e_{\mu\,a}e_{\mu\,b} = 
  [2A\det{\!(\Q^e{}_f)}]^{-1/2}\left\{\Q^c{}_a\Q_{bc}-\Q^c{}_c\Q_{ab}
  +\tsfrac12[(\Q^c{}_c)^2-\Q^c{}_d\Q^d{}_c]h_{ab}\right\},
\end{equation}
where the last frame-invariant form, should it be cumbersome to
diagonalize $p_{ab}$, is more useful than the relation
\refeq{ninvert}.

The linear relation \refeq{linrel} and Einstein's equations directly
imply that the spacetime Ricci scalar $R$ should take the constant
value $4\kappa B$. Since we must have $A>0$ for the equation of state
to be physical, three inequalities are also imposed from
\begin{equation}
  \rho-p_\mu-2B = 2A(n_{\mu+1}n_{\mu+2})^2 > 0 \quad \mbox{for all $\mu$}, 
\end{equation}
which is the same as saying that $\Q_{ab}$ should be positive
definite. All $n_\mu^{\;2}$ can subsequently be chosen to be positive,
which is the last condition to be satisfied. Hence we have a very
straightforward recipe for obtaining stationary rigid motion exact
solutions to Einstein's equation with the SUREOS.  We simply have to
look for stationary spacetimes described by $h_{ab}$, $f$ and $\omega$
as above, for which the four-dimensional Ricci scalar
\begin{equation}
  R = {}^{(3)}\!R - 2f^{-1}D^aD_af + \tsfrac12 f^{-4}D^a\omega D_a\omega 
\end{equation}
is constant. This constant should then be identified with $4\kappa B$
and should hence be nonnegative if we want the equation of state
parameter $B$ to be. If, in addition, $\Q_{ab}$ is positive definite,
we have a solution with respect to the material space metric
\refeq{matmet}. The condition $R=4\kappa B$ may clearly be written as
\begin{equation}\lbeq{DDf}
  D^aD_af = [\tsfrac12{}^{(3)}\!R-2\kappa B]f+\tsfrac14f^{-3}D^a\omega D_a\omega. 
\end{equation}
It would be very interesting to try to find twisting solutions,
especially axisymmetric solutions that could describe rotating stellar
models. However, we shall here focus on the simpler static case,
$D_a\omega = 0$, for which eq.\ \refeq{DDf} is linear and homogeneous
in $f$. In this case eq.\ \refeq{DDf} is obviously particularly simple
when the spatial Ricci scalar ${}^{(3)}\!R$ is constant, which
according to eq.\ \refeq{rho3d} corresponds to setting the energy
density $\rho$ constant. Morever, since ${}^{(3)}\!R$ and $\rho$ have
the same sign we should have ${}^{(3)}\!R > 0$ in order to obtain a
physically reasonable solution. The simplest case to consider is then
to take $h_{ab}$ to be locally isometric to the metric of a
three-sphere of some radius $r_0$, giving ${}^{(3)}\!R = 2\kappa\rho =
6/r_0^{\;2}$. Choosing coordinates such that $h_{ab}$ takes the form
\begin{equation}
  dl^2 = r_0^{\;2}\left[dx^2+\sin^2\!x\,(d\theta^2+\sin^2\!\theta\,d\phi^2)\right]
\end{equation}
and expanding $f$ according to 
\begin{equation}\lbeq{fexpand}
  f = \frac1{\sin x}\sum_{l=0}^\infty\sum_{m=-l}^lc_{lm}F_l(x)\,Y_{lm}(\theta,\phi),
\end{equation}
we obtain the radial equation 
\begin{equation}\lbeq{Fradial}
  \left[-\frac{d^2}{dx^2}+\frac{l(l+1)}{\sin^2\!x} + k^2\right]F_l = 0, 
\end{equation}
where
\begin{equation}\lbeq{kB}
  k^2 = 2(1-\kappa r_0^{\;2}B)
\end{equation}
Whether the expansion \refeq{fexpand} is appropriate to make depends
on whether or not the resulting solution can be matched to an exterior
asymptotically flat vacuum solution, or perhaps to some other matter
solution of interest. This will be a nontrivial task to analyse,
except for the spherically symmetric case; $c_{lm}=0$ for $l>0$, which
we focus on in section \ref{sec:exactfreeB}. 

\subsection{Conformally rescaled formulation}
When dealing with a four-dimensional stationary metric it is often
useful to work with the variables $\Psi = f^2$ (minus the norm of the
Killing vector $t^a$), $\omega$ and the conformally rescaled spatial
metric $\tilde{h}_{ab} = \Psi h_{ab}$, rather than $f$, $\omega$ and
$h_{ab}$. Making this reformulation here, we find that eq.\
\refeq{DDf} transforms into the equation
\begin{equation}\lbeq{Ernst-like}
  \tilde{D}^a\tilde{D}_a\Psi =
\frac{3\tilde{D}^a\Psi\,\tilde{D}_a\Psi-\tilde{D}^a\omega\,\tilde{D}_a\omega}{2\Psi}
-\tilde{R}\,\Psi + 4\kappa B,
\end{equation}
which is completely written in terms of the metric $\tilde{h}_{ab}$;
$\tilde{D}_a$ is its connection, $\tilde{R}$ its Ricci scalar and
indices are raised with its inverse, \ie with $\tilde{h}^{-1\,ab	} =
\Psi^{-1}h^{ab}$. Note that $\Psi$ and $\omega$ correspond to the real
and imaginary part, respectively, of the Ernst potential $\mathcal{E}=
\Psi + i\omega$ which in vacuum satisfies the equation
\begin{equation}
  \tilde{D}^a\tilde{D}_a\mathcal{E} =
\frac{\tilde{D}^a\mathcal{E}\,\tilde{D}_a\mathcal{E}}{\re\,\mathcal{E}}. 
\end{equation}
We expect eq.\ \refeq{Ernst-like} to be a useful complement to eq.\
\refeq{DDf}, especially when looking for solutions that can be matched
to exterior vacuum solutions, the latter being completely determined
by giving the Ernst potential $\mathcal{E}$ in the important case of
axisymmetry. Again, a more detailed investigation of these matters is
postponed for future work.

%%%%%%%%%%%%%%%%%%%%%%%%%%%%%%%%%%%%%%%%%%%%%%%%%%%%%%%%%%%%%%%%%%%%%%%%%%%%%%%%%%%%%%%%%
\section{Exact SSS solutions with constant energy density}
\label{sec:exactfreeB}
%%%%%%%%%%%%%%%%%%%%%%%%%%%%%%%%%%%%%%%%%%%%%%%%%%%%%%%%%%%%%%%%%%%%%%%%%%%%%%%%%%%%%%%%%
In the spherically symmetric case, the ansatz that the spatial metric
$h_{ab}$ be a three-sphere metric implies that the spacetime metric
can be written as
\begin{equation}
  ds^2 = \left(\frac{F}{\sin x}\right)^{\!2}dt^2 + 
  r_0^{\;2}\left[dx^2+\sin^2\!x\,(d\theta^2+\sin^2\!\theta\,d\phi^2)\right], 
  \quad r_0^{\;2} = \frac{2-k^2}{2\kappa B}, 
\end{equation}
where $F$ should satisfy eq.\ \refeq{Fradial} with $l=0$;
\begin{equation}
  F'' = k^2F. 
\end{equation}
We shall only discuss the solution that corresponds to a regular
centre model, which reads
\begin{equation}
  F = \alpha\sinh{kx}, 
\end{equation}
where $\alpha$ is a constant to be determined by the matching at the
surface of the star. The energy density and principal (radial and
tangential) pressures are given by
\begin{align}
  \rho &= \frac{6B}{2-k^2} =: \rho_0 \\
  p_r &= \frac{-(1+\cos^2\!x)\sinh{kx}+k\sin{2x}\cosh{kx}}{3\sin^2\!x\sinh{kx}}\,\rho_0 \\
  p_t &= \frac{(\cos^2\!x+k^2\sin^2\!x)\sinh{kx}-\tsfrac12k\sin{2x}\cosh{kx}}{3\sin^2\!x\sinh{kx}}\,\rho_0,
\end{align}
whereas the dimensionless quotient $m/r$ between the standard mass
function $m$ and the Schwarzschild radius $r=r_0\sin x$ takes the
simple form
\begin{equation}
  \frac{m}{r} = \tsfrac12\sin^2\!x. 
\end{equation}
Since we are dealing with a regular centre solution, the radial and
tangential pressure coincide at $x=0$. The value of this isotropic
central pressure is given by
\begin{equation}\lbeq{pc}
  p_c = \tsfrac19(2k^2-1)\rho_0 = \frac23\frac{2k^2-1}{2-k^2}\,B. 
\end{equation}
No loss of generality is implied by assuming that $k$ is non-negative
(if real) so for positive $B$ the physical range of $k$ is 
\begin{equation}\lbeq{krange}
  \frac1{\sqrt2}<k<\sqrt2,
\end{equation}
where the lower and upper limits correspond to zero and infinite
central pressure, respectively. The $k=\sqrt2$ solution is of course
singular at the centre if $B\neq 0$, but regularity can be restored if
$B$ is set to zero, in which case the constant $r_0$ becomes a scaling
parameter which is trivial in the sense that it has no relation to any
scale in the equation of state.

\begin{figure}[ht]
  \centering \begin{minipage}[t]{0.78\linewidth} 
   \centering
    \includegraphics[width=0.95\textwidth, angle=0]{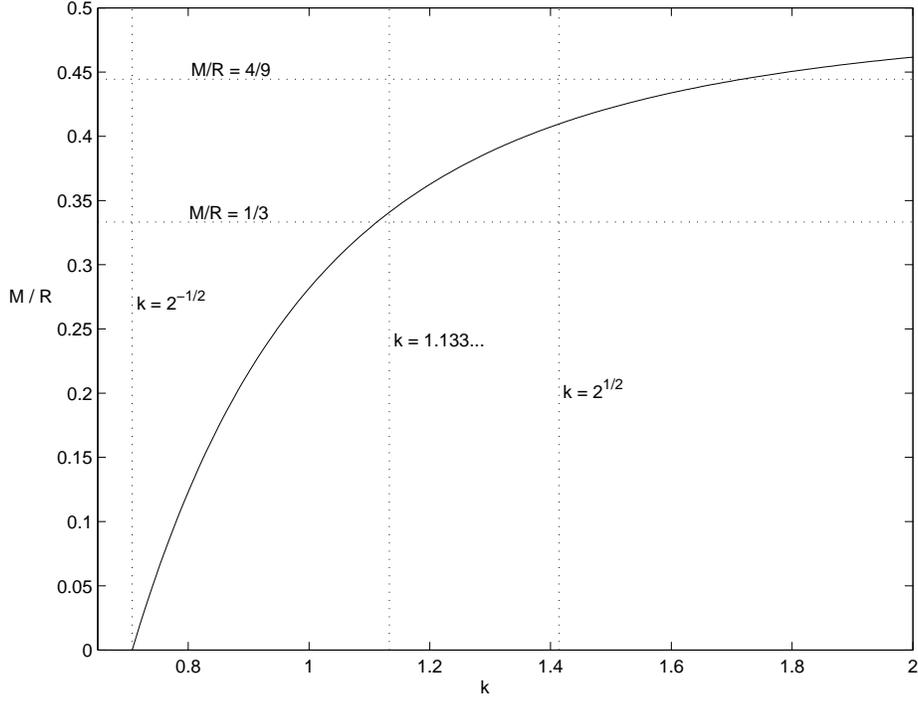} 
    \caption{The compactness $M/R$ for the SUREOS constant energy solution, plotted 
      as a function of the dimensionless parameter $k$. Although the
      discussion in this paper is restricted to non-negative values of
      $B$, which requires that $k\leq\sqrt2$, the plot has been
      continued into the negative $B$ sector $k>\sqrt2$ to display
      that the compactness then exceeds the perfect fluid Buchdahl
      limit value $M/R = 4/9$ for sufficiently large $k$. In fact, as
      $k$ is taken to infinity the curve asymptotically approaches the
      horizon value $M/R = 1/2$. Such extreme models having a surface
      arbitrarily close to the horizon have previously been
      investigated for other matter types, for charged perfect fluids
      by de Felice et al.\cite{fyf:chargedI,fsy:chargedII}. In the
      positive $B$ sector $1/\sqrt2<k<\sqrt2$ the solutions are stable
      up to $k\approx 1.133$, implying that the limiting compactness
      for the stable branch is $M/R\approx 0.3408$.  Notably this
      value is slightly larger than $1/3$ and hence there are stable
      solutions that are ultracompact in the sense that their surfaces
      are inside $R=3M$ at which the exterior Schwarzschild solution
      has closed lightlike geodesics.  At $k=\sqrt2$ (infinite central
      pressure when $B\neq 0$) the compactness is $M/R\approx
      0.4096$.}
   \label{fig:compactness} 
   \end{minipage}
\end{figure}

It is worth noting that it is possible to let $k$ exceed $\sqrt2$ if
allowing $B$ to take negative values, but this possibility will not be
pursued here (but see fig. \ref{fig:compactness}).

According to eq.\ \refeq{ninvert} the radial and tangential linear
particle densities are given by
\begin{align}\lbeq{nr4}
  n_r^{\;4} &=  \frac{(\rho-p_t-2B)^2}{2A(\rho-p_r-2B)} \\ \lbeq{nt4}
  n_t^{\;4} &=  \frac{\rho-p_r-2B}{2A}, 
\end{align}
in terms of which the pull-back of the material space metric is
\begin{equation}\lbeq{materialmetric}
  k_{ab} = n_r^{\;2}r_ar_b + n_t^{\;2}t_{ab}, 
\end{equation}
where
\begin{equation}
  r_a = r_0\,(dx)_a, \quad 
  t_{ab} = r_0^{\;2}\sin^2{\!x}\,(d\theta^2+\sin^2{\!\theta}\,d\phi^2)_{ab}.  
\end{equation}
Obviously we could write down the metric \refeq{materialmetric}
explicitly in the coordinates $(x,\theta,\phi)$, but refrain from
doing so since the result is not very illuminating. Note, however,
that $k_{ab}$ always has a regular centre whenever the spacetime
metric $g_{ab}$ does.

From eqs.\ \refeq{transv2}, \refeq{nr4} and
\refeq{nt4} we can also calculate the three generally distinct
transversal elastic wave speeds as
\begin{align}
  v_{r\perp t}^{\;2} &= \frac{n_t^{\;2}}{n_t^{\;2}+n_r^{\;2}}  \\
  v_{t\perp r}^{\;2} &= \frac{n_t^{\;2}}{n_t^{\;2}+n_t^{\;2}} \equiv \frac12 \\
  v_{t\perp t}^{\;2} &= \frac{n_r^{\;2}}{n_r^{\;2}+n_t^{\;2}}.
\end{align}

The surface of the model will be at a radius for which the radial
pressure vanishes, allowing for a matching to the vacuum Schwarzschild
solution to be made. For the allowed values of $k$ such a surface
always exists at a finite distance from the centre, although we have
no closed analytic form for the solution to the equation $p_r = 0$.
However, the constant $\alpha$ can be determined by the matching and
is given by
\begin{equation}
  \alpha = \tsfrac12\sin{2x_\mathrm{s}}\sinh{kx_\mathrm{s}},
\end{equation}
where the subscript $\mathrm{s}$ refers to evaluation on the surface. 

\begin{figure}[ht]
 \centering \begin{minipage}[t]{0.78\linewidth}
  \centering
  \includegraphics[width=0.95\textwidth, angle=0]{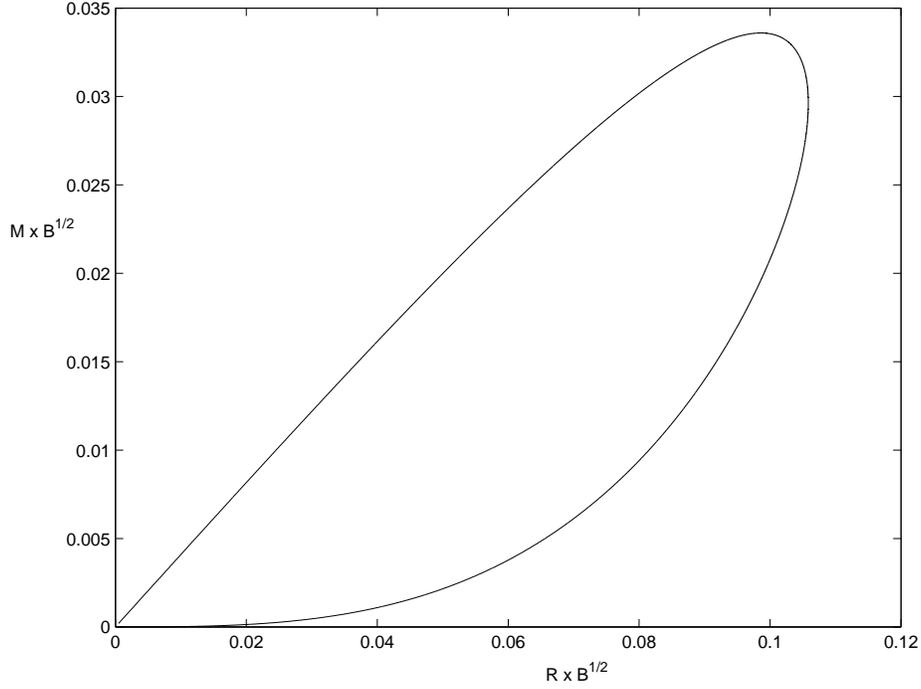}
   \caption{Mass-radius curve for the SUREOS constant energy density solution 
     with $B$ at a fixed positive value. The curve is parametrized by
     the dimensionless parameter $k$ in the range $1/\sqrt2<k<\sqrt2$
     or equivalently by the central pressure \mbox{$p_c =
       \frac23B(2k^2-1)/(2-k^2)$} taking all positive values.  The
     direction of increasing $k$ and $p_c$ is the counterclockwise
     one. The maximum of the curve is at $k\approx 1.133$ and
     separates this one-parameter family of solutions into a stable
     branch with $1/\sqrt2<k\lesssim 1.133$ and a branch with
     $1.133\lesssim k<\sqrt2$ having one unstable radial mode. A
     peculiar feature of this mass-radius curve is that it not only
     starts out from the origin $M=R=0$ at zero central pressure but
     also returns there in the limit of infinite central pressure.}
 \label{fig:MR}
 \end{minipage}
\end{figure}

\begin{figure}[ht]
  \centering \begin{minipage}[t]{0.8\linewidth} 
   \centering
    \includegraphics[width=0.95\textwidth, angle=0]{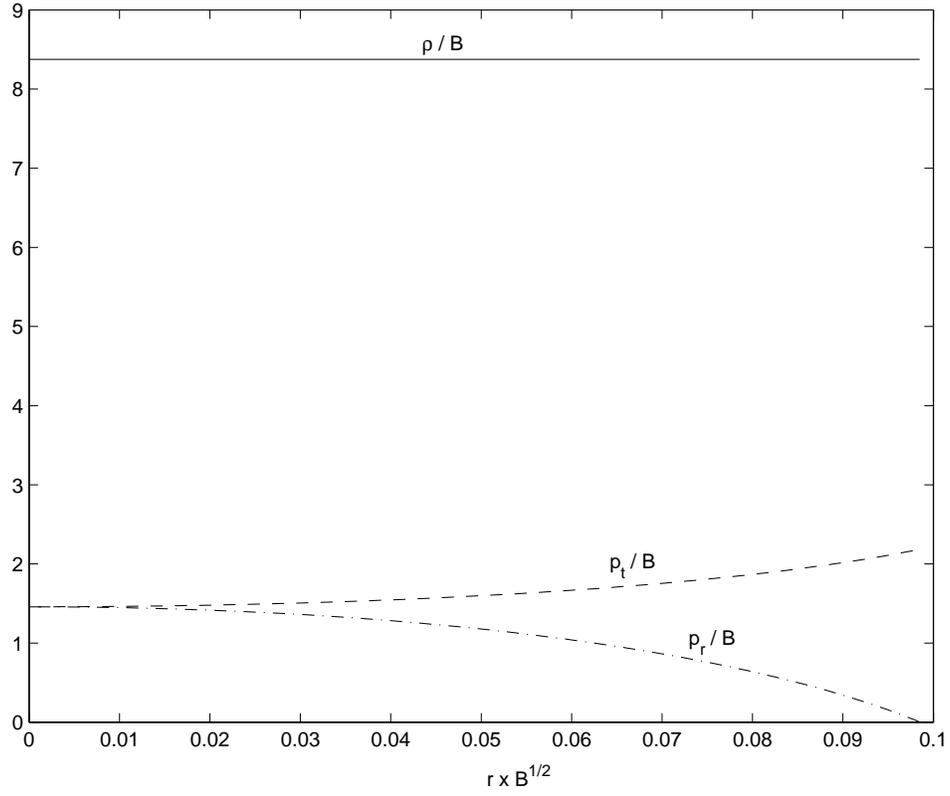} 
      \caption{Energy density
        and pressure profiles for the marginally stable model with
        $k\approx 1.133$. The radius of this model is $R\approx
        0.09857\times B^{-1/2}$. The radial (tangential) pressure is
        monotonically decreasing (increasing) for other values of $k$
        as well.}
     \label{fig:prof_crit} 
     \end{minipage} 
\end{figure}

\begin{figure}[hb]
  \centering \begin{minipage}[t]{0.8\linewidth} 
   \centering
    \includegraphics[width=0.95\textwidth, angle=0]{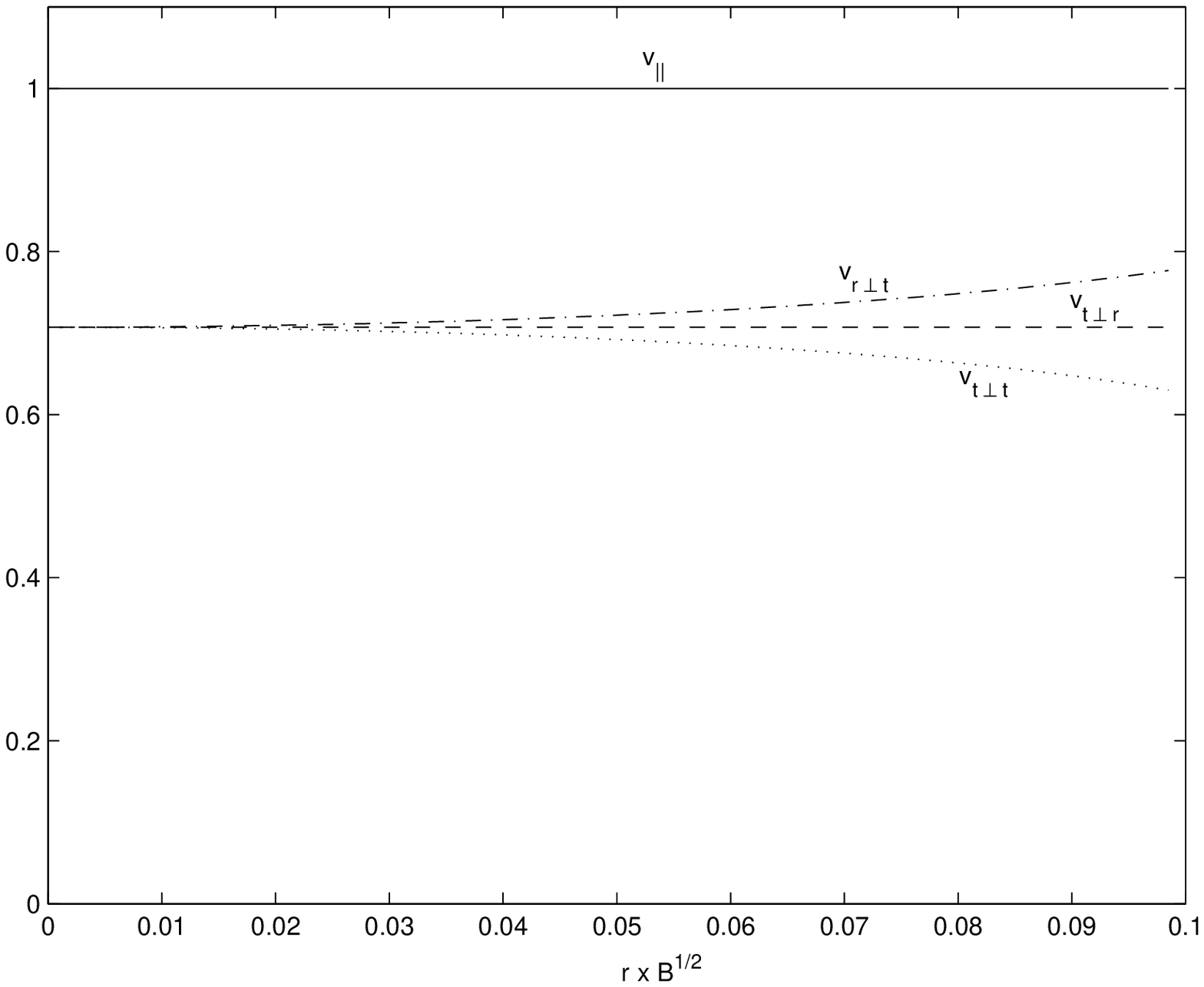} 
      \caption{Profiles of the speeds of elastic wave propagation for the marginally 
        stable model with $k\approx 1.133$. While the single
        longitudinal wave speed $v_{||}$ is always unity (speed of
        light) the principal transversal wave speeds $v_{r\perp t}$,
        $v_{t\perp r}$ and $v_{t\perp t}$ start out from the unsheared
        value $\sqrt2$ at the centre and diverges outwards as the
        pressure anisotropy increases. For smaller (larger) values of
        $k$, and thus of the central pressure, the profiles are
        qualitatively similar but the divergence is weaker (stronger).
        The transversal speed $v_{t\perp r}$ is however always exactly
        equal to $\sqrt2$.}
     \label{fig:waves_crit} 
     \end{minipage} 
\end{figure}

The further properties of this constant energy density family of
solutions are presented in figures \ref{fig:MR}-\ref{fig:freqs} and
their captions.

%%%%%%%%%%%%%%%%%%%%%%%%%%%%%%%%%%%%%%%%%%%%%%%%%%%%%%%%%%%%%%%%%%%%%%%%%%
%%%%%%%%%%%%%%%%%%%%%%%%%%%%%%%%%%%%%%%%%%%%%%%%%%%%%%%%%%%%%%%%%%%%%%%%%%%
%\newpage
%%%%%%%%%%%%%%%%%%%%%%%%%%%%%%%%%%%%%%%%%%%%%%%%%%%%%%%%%%%%%%%%%%%%%%%%%%%
%%%%%%%%%%%%%%%%%%%%%%%%%%%%%%%%%%%%%%%%%%%%%%%%%%%%%%%%%%%%%%%%%%%%%%%%%%%%%%%
%\newpage
%%%%%%%%%%%%%%%%%%%%%%%%%%%%%%%%%%%%%%%%%%%%%%%%%%%%%%%%%%%%%%%%%%%%%%%%%%%%%%%
%%%%%%%%%%%%%%%%%%%%%%%%%%%%%%%%%%%%%%%%%%%%%%%%%%%%%%%%%%%%%%%%%%%%%%%%%%%%%%%
%\newpage
%%%%%%%%%%%%%%%%%%%%%%%%%%%%%%%%%%%%%%%%%%%%%%%%%%%%%%%%%%%%%%%%%%%%%%%%%%%%%%%
%%%%%%%%%%%%%%%%%%%%%%%%%%%%%%%%%%%%%%%%%%%%%%%%%%%%%%%%%%%%%%%%%%%%%%%%%%

%%%%%%%%%%%%%%%%%%%%%%%%%%%%%%%%%%%%%%%%%%%%%%%%%%%%%%%%%%%%%%%%%%%%%%%%%%
\subsection{Stability against radial perturbations}\label{sec:stability}
\begin{figure}[hb]
  \centering \begin{minipage}[t]{0.8\linewidth} 
   \centering
    \includegraphics[width=0.95\textwidth, angle=0]{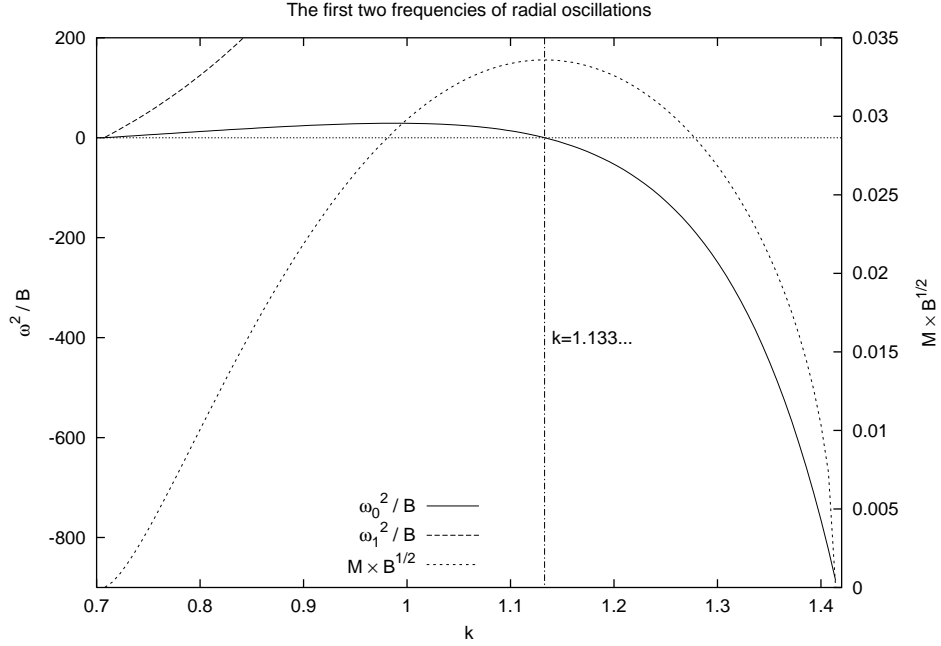} 
    \caption{The first two frequencies of radial oscillations for the 
      family of constant energy density models with $B>0$. The squared
      frequencies as well as the stellar mass $M$ are plotted as
      functions of the dimensionless parameter $k$, displaying clearly
      that $\omega_0^{\;2}$ turns negative precisely at the maximum
      mass model having $k\approx 1.133$. Although partially cut out
      from the plot, the square of the first overtone frequency
      $\omega_1$ remains positive up to the infinite central pressure
      limit $k=\sqrt2$.}
   \label{fig:freqs} 
   \end{minipage}
\end{figure}

%%%%%%%%%%%%%%%%%%%%%%%%%%%%%%%%%%%%%%%%%%%%%%%%%%%%%%%%%%%%%%%%%%%%%%%%%%
According to the results of paper II, the radial perturbations for the
constant energy density solutions are governed by the first order
system
\begin{align}\lbeq{dzetadx}
  \frac{d\zeta}{dx} &= r_0\cos x\,[Y\zeta+P^{-1}\eta] \\ \lbeq{detadx}
  \frac{d\eta}{dx} &= -r_0\cos x\,[(Q+\omega^2W)\zeta+Y\eta],
\end{align}
where the indendent variables $\zeta$ and $\eta$ are closely related
to the lagrangian perturbations of the Schwarzschild radius and the
radial pressure, respectively. Moreover $r_0 = \sqrt{(2-k^2)/(2\kappa
  B)}$ and
\begin{align}
  P &= \frac{2\alpha^3}{\kappa r_0^{\;4}}
  \frac{(-\cos{2x}\sinh{kx}+\tsfrac12 k\sin{2x}\cosh{kx})\sinh^2{\!kx}}{\sin^7{\!x}\cos{x}} \\
  Y &= -\frac1{r_0}\frac{[(2+k^2)\sin^2{\!x}+1]\sinh{kx}
  -\tsfrac12 k\sin{2x}\cosh{kx}}{(\cos{2x}\sinh{kx}-\tsfrac12 k\sin{2x}\cosh{kx})\sin{x}} \\
  W &= P\alpha^{-2}\frac{\tan^2\!x}{\sinh^2{\!kx}} \\
  Q &= Q_1 + Q_2 \\
  Q_1 &= P\frac4{r_0^{\;2}}\frac{k^2\sin^2{\!x}\cosh^2{\!kx}
   -\cos{2x}\sinh^2{\!kx}}{\sin^2{\!2x}\sinh^2{\!kx}} \\
  Q_2 &= P(Z+Y^2) \\
  Z &= \frac1{r_0^{\;2}}\frac{3\cos{x}\left[(2\cos^2{\!x}+5)\sinh^2{\!kx}
   +k^2\sin^2{\!x}\,(5\cosh^2{kx}-3)\right] - k\sin{x}\,(1+\tsfrac{25}2\cos^2{\!x}
   +k^2\sin^2{\!x})\sinh{2kx}}{(\cos{2x}\sinh{kx}
   -\tsfrac12 k\sin{2x}\cosh{kx})\sin^2{\!x}\cos{x}\sinh{kx}}. 
\end{align}
A regular solution to the system \refeq{dzetadx} and \refeq{detadx}
should leave the centre according to
\begin{align}
  \zeta &= Cr_0^{\;3}x^3 + O(x^5) \\
  \eta &= CP_0(3-Y_0) + O(x^2), 
\end{align}
where $C$ is an arbitrary constant and where $P_0$ and $Y_0$, as
defined in paper II, are given by
\begin{align}
  P_0 &= \frac23\frac{\alpha^3k^3(k^2+4)}{\kappa r_0^{\;2}} \\
  Y_0 &= 2. 
\end{align}
At the surface of the star the lagrangian perturbation of the radial
pressure should vanish, implying
\begin{equation}
  \eta|_{x=x_{\mathrm{s}}} = 0. 
\end{equation}
We have numerically determined the first two frequencies $\omega_0$
and $\omega_1$ of radial oscillations for $B>0$ and found that
$\omega_0^{\;2}$ changes sign from positive to negative at $k\approx
1.133$, which to numerical precision\footnote{We use standard methods
as described in \cite{nr}. More precisely we find the frequencies
using the shooting method employing a fourth-fifth order adaptive step
size Runge-Kutta ODE-solver. The change of sign of the eigenvalue
$\omega_0^{\;2}$ coinsides with the maximum of the mass to within more
than six significant digits in the mass.}
coincides with the maximum mass model. As expected,
$\omega_1^{\,2}$ is positive for all allowed values of $k$.  The
results are presented in figure \ref{fig:freqs}.
%\begin{align}
%  P_0 &= \frac23\frac{\alpha^3k^3(k^2+4)}{\kappa r_0^{\;2}} \\
%  Y_0 &= 2 \\
%  W_0 &= \frac23\frac{\alpha k(k^2+4)}{\kappa r_0^{\;2}} \\
%  Q_0 &= -P_0Y_0(3-Y_0) = -2P_0. 
%\end{align}
%%%%%%%%%%%%%%%%%%%%%%%%%%%%%%%%%%%%%%%%%%%%%%%%%%%%%%%%%%%%%%%%%%%%%%%%%%%
%%%%%%%%%%%%%%%%%%%%%%%%%%%%%%%%%%%%%%%%%%%%%%%%%%%%%%%%%%%%%%%%%%%%%%%%%%
\clearpage

\section{A solution with nonconstant energy density}\label{sec:exactzeroB}
%%%%%%%%%%%%%%%%%%%%%%%%%%%%%%%%%%%%%%%%%%%%%%%%%%%%%%%%%%%%%%%%%%%%%%%%%%
The constant energy density of the family of solutions presented in
the previous section comes about because the unsheared, \ie
compressional, energy density $\check\rho$ decreases outwards at
exactly the same rate as the shearing energy density $\sigma =
\rho-\check\rho$ inreases. Here we will briefly present a different
SUREOS solution which is more realistic in the sense that the energy
density as well as both principal pressures are monotonically outwards
decreasing, just to illustrate that such solutions do exist. The
solution was found for $B=0$ only and although we have made some
attempts to generalise it to nonzero $B$ we have not been able to do
so.

\begin{figure}[ht]
  \centering \begin{minipage}[t]{0.8\linewidth} \centering
    \includegraphics[width=0.95\textwidth,angle=0]{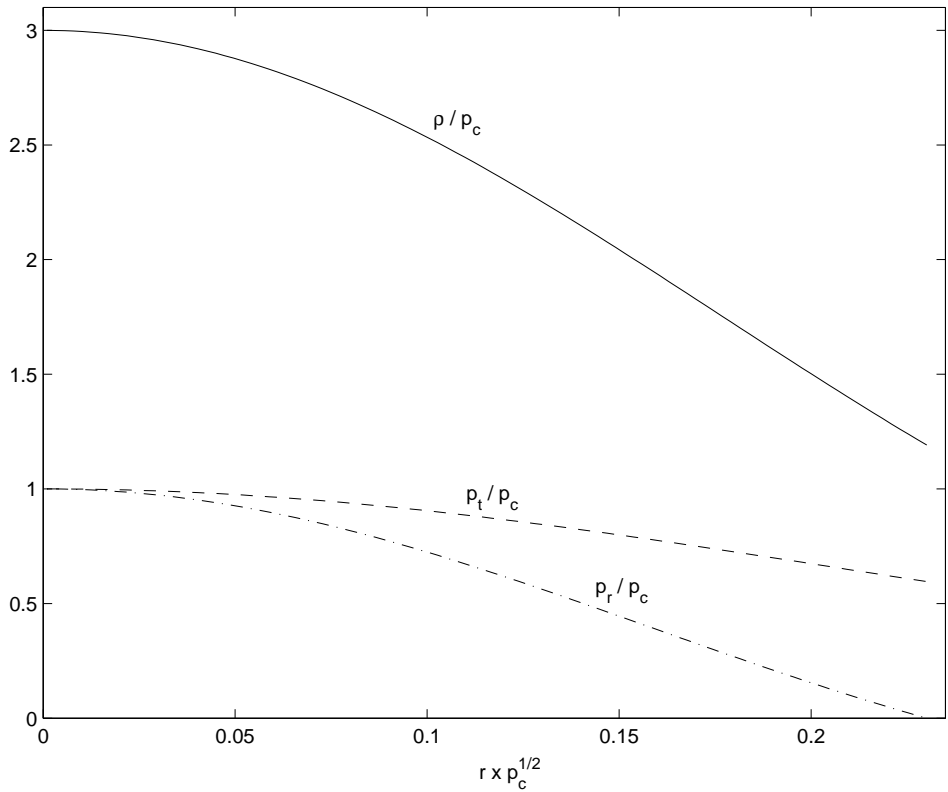} 
    \caption{Energy density and pressure profiles for the $B=0$ solution with 
     outwards decreasing energy density and eigenpressures.}
    \label{fig:nonconst} 
    \end{minipage} 
\end{figure}

In Schwarzschild coordinates the spacetime metric for the solution has
the simple analytic form
\begin{equation}
  ds^2 = \sinc^{\!-2}{\!(r/r_0)}\,(-\alpha^2dt^2+dr^2) + r^2d\Omega^2, \quad 
  \sinc{x} = \frac{\sin{x}}{x}. 
\end{equation}
To match the solution to a Schwarzschild exterior without rescaling
the time coordinate, the constant $\alpha$ should be chosen as
\begin{equation}
  \alpha = \sinc^2{\!(R/r_0)},
\end{equation}
where $R$ is the radius of the star, \ie $r=R$ is the surface of
vanishing radial pressure, to be determined below. Regularity at the
centre is clearly guaranteed by the fact that $\sinc x = 1$ at $x =
0$. The energy density and principal pressures are given by
\begin{align}
  \kappa r^2\rho &= 1 + \sinc^2{\!(r/r_0)} - 2\sinc{\!(r/r_0)}\cos{\!(r/r_0)} \\
  \kappa r^2 p_r &= -1 + 3\sinc^2{\!(r/r_0)} - 2\sinc{\!(r/r_0)}\cos{\!(r/r_0)} \\
  \kappa r^2 p_t &= \frac{2m}{r} = 1 - \sinc^2{\!(r/r_0)} \\
\end{align}
\begin{equation}
  \rho_c = 3p_c = \frac1{\kappa r_0^{\;2}}.
\end{equation}
The profiles of these quantities are plotted in figure
\ref{fig:nonconst}. The radial pressure drops to zero very close to
$r=2r_0$, or more precisely
\begin{equation}
  R \approx 1.9979\,r_0,
\end{equation}
implying $\alpha\approx 0.2075$ and $R/M\approx2.5237$.

A similar analysis to that of section \ref{sec:stability} reveals that
the fundamental radial mode is unstable with squared frequency
$\omega_0^{\;2}\approx -0.07017\,r_0^{-2}$ which for a typical neutron
star with a radius of the order of 10 km corresponds to an e-folding
time of about 0.4 ms. Clearly then, this solution is not very
interesting on its own right, but if it could be generalised to a
family with $B$ being a free parameter we expect there to be a stable
branch of solutions. 

%%%%%%%%%%%%%%%%%%%%%%%%%%%%%%%%%%%%%%%%%%%%%%%%%%%%%%%%%%%%%%%%%%%%%%%
\section{Discussion and outlook}
Exact solutions to the Einstein equations can very rarely be used to
completely replace numerically obtained solutions, with a few
important exceptions such as the Kerr-Newman family of black hole
solutions. The reason is of course that it is almost always impossible
to find an exact solution that obeys all restrictions that are imposed
by requiring that the solution be astrophysically realistic. However,
if an exact solution satisfies at least some basic criteria for
physicality, that solution has the potential of being a useful tool
for gaining qualitative insights. Furthermore it can be very useful as
a test background model when doing numerical studies of various types
of perturbations.  The constant energy density family of exact SSS
solutions with stiff ultrarigid equation of state does, in our
opinion, satisfy the relevant list of basic criteria that renders it
useful in these respects. 
In fact we have already used it in the present paper to test that our
analysis and numerical codes for radial oscillations of spherically
symmetric elastic matter models are behaving as expected; Clearly it
cannot be a coincidence that we find the zeroth mode turning unstable
for the maximum mass member of the family, to numerical
precision.
So far this had only been
tested for the numerically integrated models presented in paper I,
which all have moderate pressure anisotropies, but now we know that it
also works as expected for models with more extreme anisotropies.

Since the presented method of generating SUREOS solutions does not
rely upon spherical symmetry but only on stationarity and rigid
motion, it would be very nice if one could also apply it to finding a
rigidly rotating solution family. 
%Although the equation of state is
%rather extreme, it is nevertheless always microstable and causal and
%hence, in our view, more physical than the equation of state
%$\rho+3p=\rho_0$ for Wahlquist's rotating perfect fluid
%solution\cite{wahlquist:rotating} whose squared speed of sound
%$dp/d\rho$ is negative and hence fails to be microstable.
%%%%korr av syftningsfel$$$$
Although the equation of state is rather extreme, it is nevertheless
always microstable and causal and hence, in our view, more physical
than the equation of state $\rho+3p=\rho_0$ for Wahlquist's rotating
perfect fluid solution\cite{wahlquist:rotating}. The latter equation
of state fails to be microstable since the squared speed of sound
$dp/d\rho$ is obviously negative.

\section*{Appendix}
Since the principal longitudinal speed is equal to the speed of light,
it is of interest to find out whether this speed can become
superluminal when the propagation vector $\nu^a$ is perturbed around
the principal unit vector $e_\mu^{\;a}$. For the SUREOS the Fresnel
tensor can be written in the simple form
\begin{equation}
  Q^{ab} = \rho h^{ab}+p^{ab} - L^{ab},
\end{equation}
where $L^{ab}$ is the tensor
\begin{equation}
  L^{ab} = \sum_{\sigma=1}^3(\rho-p_\sigma-2B)Z_\sigma^{\;a} Z_\sigma^{\;b} = 
  2A\sum_{\sigma=1}^3n_{\sigma+1}^{\;2}n_{\sigma+2}^{\;2}Z_\sigma^{\;a} Z_\sigma^{\;b}, 
  \quad 
  Z_\sigma^{\;a} = \epsilon^{abc}e_{\sigma b}\,\nu_c =\nu_{\sigma+1}\, 
  e_{\sigma+2}^{\;a}-\nu_{\sigma+2}\,e_{\sigma+1}^{\;a}.
\end{equation}
Thus the characteristic equation, which given a unit propagation
vector $\nu^a = \sum_{\sigma=1}^3\nu_\sigma e_\sigma^{\;a}$ determines
three propagation speeds $v$ and polarization vectors $\iota^a$, can
in this case be rewritten in the convenient form
\begin{equation}\lbeq{charsimple}
  \left[(1-v^2)(\rho h^{ab}+p^{ab}) + L^{ab}\right]\iota_b = 0.
\end{equation}
Since all three vectors $Z_\sigma^{\;a}$ are orthogonal to $\nu^a$, so
is the tensor $L^{ab}$ which means that the characteristic equation
always has the solution
\begin{align}
  v^2 &= 1 \\
  \iota_a &= \nu_a
\end{align}
Since this result holds regardless of the choice of propagation vector
$\nu^a$, we have proved the claim that the equation of state allows
for a purely longitudinally polarized wave mode in \emph{all}
directions, \ie not only principal directions which is the case
generically. Moreover, the speed of these waves is direction
independent and equal to the speed of light.  Let us now consider
$L^{ab}$ as a tensor on the two-space orthogonal to $\nu^a$ (and, of
course, to $u^a$). As such, its determinant (or, rather, the
determinant of the mixed tensor $L^a{}_b$) can be calculated
according to
\begin{equation}
  \tsfrac12 L^{ab}L^{cd}\epsilon_{ac}\epsilon_{bd}, 
  \quad \epsilon_{ab} = \epsilon_{abc}\nu^c. 
\end{equation}
The result is 
\begin{equation}
  \tsfrac12 L^{ab}L^{cd}\epsilon_{ac}\epsilon_{bd} = 
  \sum_{\sigma=1}^3(\rho-p_{\sigma+1}-2B)(\rho-p_{\sigma+2}-2B)\nu_\sigma^{\;2} 
  = 4A^2n^2\sum_{\sigma=1}^3n_\sigma^{\;2}\nu_{\sigma}^{\;2}. 
\end{equation}
Clearly, this scalar can only be zero if at least one linear particle
density vanishes which only happens if the material space mapping is
degenerate which is disallowed. Now, with $L^{ab}$ being orthogonal to
$\nu^a$ and of rank two, it follows that the only solution to the
characteristic equation \refeq{charsimple} with $v^2=1$ is the
longitudinally polarized solution. For, if $\iota_a$ was to have a
nonvanishing projection orthogonal to $\nu^a$, that projection would
have to be annihilated by $L^{ab}$ which is impossible by the
nondegeneracy property of the latter. As a corollary we thus have
proved that the remaining two polarization modes always have
propagation speeds less than unity.  This follows from the facts that
they should vary continuously with the propagation direction and that
they are strictly less than unity for propagation in principal
directions, in which case the polarization is purely transversal.

%\bibliographystyle{../../bib/prsty}
%\bibliography{../../max}

%%%%%%%%%%%%%%%%%%%%%%%%%%%%%%%%%%%%%%%%%%%%%%%%%%%%%%%%%%%%%%%%%%%%%%

%%%%%%%%%%%%%%%%%%%%%%%%% 

\end{document}